**Partial Collapse and Ensemble Invariance under Continuous Quantum Measurement**

**Shalender Singh**, **Santosh Kumar**
*PolariQon Inc., Palo Alto, California, USA*
*(December 2025)*

**Abstract**

Wavefunction collapse is commonly associated with unavoidable physical disturbance of the measured system. Here we show that in driven–dissipative quantum systems, continuous measurement can induce strong trajectory-level collapse while leaving the ensemble-averaged steady state strictly invariant. We identify measurement-invariant steady states whose unconditional density matrix remains unchanged under continuous monitoring, despite pronounced measurement-induced localization in conditioned quantum trajectories. This separation between trajectory-level collapse and ensemble invariance defines a regime of *partial collapse*, in which measurement-induced localization is continuously counteracted by dissipative dynamics. We derive a necessary and sufficient condition for steady-state invariance under continuous measurement and identify Liouvillian symmetry as a concrete dynamical mechanism enforcing it. Our results clarify the distinction between conditional collapse and physical disturbance in open quantum systems and provide a framework for non-invasive continuous monitoring in driven–dissipative settings.

## I. INTRODUCTION

Quantum measurement is traditionally associated with unavoidable disturbance of the measured system, culminating in wavefunction collapse [1–3]. In closed systems, this intuition is formalized by projective measurement theory, where information gain and state disturbance are fundamentally inseparable [4]. In such settings, protocols that avoid disturbance—such as standard quantum non-demolition (QND) measurements—are possible only when the measured observable is conserved, commuting with both the system Hamiltonian and the measurement interaction.

In contrast, many contemporary quantum platforms operate far from equilibrium, sustained by continuous drive and dissipation and naturally described by open quantum system dynamics [5–7]. These systems generally do not admit conserved observables of the kind required for textbook non-demolition measurement. As a result, continuous monitoring in driven–dissipative systems is generically expected to modify the physical steady state.

Continuous quantum measurement provides a particularly sharp lens on this distinction [8–11]. When a system is monitored weakly and continuously, the evolution of the quantum state depends on whether one conditions on the measurement record. At the level of individual trajectories, continuous measurement induces stochastic localization—often interpreted as gradual collapse—while the ensemble-averaged state evolves deterministically under a modified master equation [12–15]. Such trajectory-resolved dynamics have been observed and reconstructed experimentally in platforms including superconducting circuits and cavity QED, where continuous monitoring and long-time steady states are simultaneously accessible [22–24]. Such protocols are now routinely implemented in experimental platforms including superconducting circuits, cavity QED, and trapped ions, where both individual trajectories and long-time steady states are directly accessible. This separation between conditional and unconditional dynamics raises a fundamental question: must continuous measurement necessarily disturb the physical steady state of an open, driven–dissipative quantum system?
In this work, we show that the answer is no. We focus on a well-defined and experimentally relevant class of driven–dissipative systems whose unmeasured dynamics are governed by purely incoherent processes that are covariant under phase rotations generated by the measured observable. While this class does not encompass all driven quantum systems, it captures a wide range of thermalizing, pumping, and loss-dominated platforms where coherent driving is weak or absent. These systems admit measurement-invariant steady states, for which the unconditional density matrix remains unchanged under continuous monitoring. Crucially, this invariance does not arise from a conserved observable or an imposed non-demolition constraint. Instead, it emerges dynamically from the interplay of continuous measurement, external drive, and dissipation. More broadly, the role of symmetries and conserved structures of Lindblad generators in constraining steady states and long-time dynamics of open quantum systems has been studied extensively in other contexts [25,26]. Here we provide a precise criterion for when such structure renders continuous measurement non-invasive at the level of the steady-state ensemble.

In such systems, measurement extracts genuine information and induces collapse at the level of conditioned trajectories, yet the physical steady-state ensemble is preserved. This separation gives rise to a regime we term *partial collapse*, in which measurement-induced localization is transient and continuously counteracted by the system's native dissipative dynamics.

The distinction between conditional (trajectory-level) dynamics and unconditional (ensemble-averaged) evolution is a foundational feature of continuous quantum measurement theory and has been well understood since the early development of the stochastic master-equation formalism [8–15]. The present work does not revisit this formal separation. Rather, it addresses a more specific and previously under-characterized question: **under what dynamical conditions does continuous measurement leave the physical steady-state ensemble strictly invariant, despite inducing strong trajectory-level localization?**

While ensemble invariance is trivial in closed-system QND measurements due to exact conservation laws, its occurrence in driven–dissipative open systems—where no such conservation exists—is not generally guaranteed and has not been systematically characterized in the continuous-measurement literature. While Liouvillian symmetries and invariant steady states have been studied extensively in open quantum systems [25,26], their role in enforcing ensemble invariance specifically under continuous measurement—and the resulting coexistence of trajectory-level collapse with steady-state preservation—has not been isolated as a distinct dynamical regime. Here we provide a precise criterion and a concrete dynamical mechanism by which such invariance can arise autonomously in open systems, and we characterize the resulting regime of transient, trajectory-level localization coexisting with strict ensemble invariance.

Our results rely only on standard open quantum system theory and do not invoke modifications of quantum mechanics. We derive a necessary and sufficient condition for steady-state invariance under continuous measurement and show that it holds for arbitrary measurement strength within the validity of the Lindblad description. This framework clarifies how measurement backaction can be dynamically neutralized at the ensemble level, even when the measurement is informationally nontrivial, and sharpens the conceptual distinction between collapse as a physical process and collapse as conditioned knowledge [16–18].

Beyond foundational interest, measurement-invariant steady states are directly relevant to continuous quantum sensing, long-time monitoring, and nonprojective measurement protocols in driven platforms [19–21]. More broadly, they establish a unifying perspective in which non-demolition measurement and driven–dissipative steady-state stabilization represent distinct physical mechanisms that can nonetheless yield equivalent operational outcomes under continuous monitoring.

**Relation to quantum non-demolition measurement**

In conventional quantum non-demolition (QND) measurement, ensemble invariance follows from a kinematic conservation law: the measured observable $c$commutes with the system Hamiltonian, $[H, c] = 0$, ensuring that measurement backaction does not alter its statistics in a closed system. In driven–dissipative settings, such Hamiltonian conservation is generally absent; however, symmetry may instead be enforced at the level of the full Liouvillian generator.

Here we show that when the unmeasured driven–dissipative dynamics are covariant under the phase rotations generated by the measured observable, the resulting steady state lies in the kernel of the measurement-induced dephasing superoperator. Continuous measurement then leaves the steady-state ensemble invariant, even though individual quantum trajectories undergo stochastic localization and dissipative transitions. Measurement-invariant steady states in driven–dissipative systems therefore constitute a **Liouvillian generalization of QND measurement**, in which symmetry is enforced dynamically by the combined action of Hamiltonian and dissipative processes rather than by Hamiltonian conservation alone.

**Novelty and scope of the present work**

In this work, we identify and characterize the conditions under which continuously measured, non-Hermitian driven–dissipative quantum systems admit measurement-invariant steady states, despite exhibiting strong trajectory-level collapse. Under these conditions, the unconditional steady-state ensemble remains unchanged by continuous monitoring, so that standard stochastic master-equation theory applies without modification at the ensemble level.

We provide a necessary and sufficient criterion for steady-state invariance, identify Liouvillian symmetry as a concrete dynamical mechanism enforcing it, and show that this mechanism enables an operationally non-invasive measurement regime without requiring conserved observables, quantum non-demolition Hamiltonians, feedback, or postselection. In this sense, the work formalizes and isolates a dynamical regime that is implicit but not systematically identified in standard treatments of continuous quantum measurement.

By reducing the analysis of continuously monitored driven–dissipative systems to an invariant steady-state ensemble whenever these conditions are satisfied, the framework converts a generally complex non-equilibrium measurement problem into one that can be analyzed and operationalized using well-established stochastic-trajectory methods.

## 2 RESULTS

### 2.1 Conditional and Unconditional Dynamics Under Continuous Measurement

We consider a driven–dissipative quantum system whose unconditional dynamics are governed by a Lindblad master equation

$$\dot{\rho} = \mathcal{L}\rho = -i[H, \rho] + \sum_k \mathcal{D}[L_k]\rho, \qquad (1)$$

where $H$is the system Hamiltonian, $L_k$are dissipative jump operators, and

$$\mathcal{D}[L]\rho = L\rho L^\dagger - \frac{1}{2}\{L^\dagger L, \rho\} \qquad (2)$$

denotes the standard Lindblad dissipator. We assume that Eq. (1) admits a unique steady state $\rho_{\mathrm{ss}}$ satisfying $\mathcal{L}\rho_{\mathrm{ss}} = 0$.

We now introduce continuous monitoring of a system observable associated with measurement operator $c$, performed with measurement rate $\Gamma_m$ and efficiency $\eta$. Conditioning on the measurement record yields a stochastic master equation (SME) for the conditioned state $\rho_c(t)$,

$$d\rho_c = (\mathcal{L} + \Gamma_m \mathcal{D}[c])\rho_c \, dt + \sqrt{\eta\Gamma_m}\, \mathcal{H}[c]\rho_c \, dW_t, \qquad (3)$$

where $dW_t$ is a Wiener increment satisfying $\mathbb{E}[dW_t] = 0$and $\mathbb{E}[dW_t^2] = dt$, and

$$\mathcal{H}[c]\rho = c\rho + \rho c^\dagger - \mathrm{Tr}[(c + c^\dagger)\rho]\rho \qquad (4)$$

encodes the measurement-induced innovation associated with information gain.

Equation (3) makes explicit the distinction between conditional and unconditional dynamics. The stochastic term proportional to $\mathcal{H}[c]$ drives localization of the conditioned state and is responsible for measurement-induced collapse at the level of individual trajectories. In contrast, averaging over all measurement records eliminates the stochastic contribution and yields a deterministic evolution for the ensemble-averaged state,

$$\mathbb{E}[d\rho_c] = (\mathcal{L} + \Gamma_m \mathcal{D}[c])\rho \, dt. \qquad (5)$$

**Information gain under continuous measurement.**
In the stochastic master equation (3), information extraction is encoded in the innovation term proportional to $\sqrt{\eta\Gamma_m}\, \mathcal{H}[c]\rho_c \, dW_t$. A natural quantitative measure of information gain is the reduction of entropy of the conditioned state relative to the ensemble,

$$\Delta S(t) = S(\rho_{\mathrm{ens}}) - \mathbb{E}[S(\rho_c(t))],$$

where $S(\rho) = -\mathrm{Tr}(\rho \log \rho)$ is the von Neumann entropy and $\rho_{\mathrm{ens}} = \mathbb{E}[\rho_c]$.
Positive $\Delta S(t)$ indicates genuine information acquisition from the measurement record.

Continuous measurement therefore modifies the unconditional dynamics solely through the additional dissipator $\Gamma_m \mathcal{D}[c]$. Whether measurement perturbs the physical steady state is thus determined entirely by the action of this term on $\rho_{\mathrm{ss}}$. This observation motivates the central question of this work: under what conditions does continuous measurement leave the steady-state ensemble invariant, despite inducing collapse in the conditioned dynamics? We address this question in the next section by identifying necessary and sufficient conditions for measurement-invariant steady states.

Physically, Eq. (5) shows that continuous measurement perturbs the steady state only through its unconditional dephasing action, implying that measurement-induced collapse at the trajectory level need not translate into disturbance of the physical ensemble.

*Remark*

Equations (3) – (5) rely entirely on standard stochastic master-equation theory; both trajectory-level dynamics and ensemble-averaged evolution are well established. The problem addressed here is to determine when this framework remains valid for continuously monitored, non-Hermitian driven–dissipative systems, where measurement adds an explicit dephasing term $\Gamma_m \mathcal{D}[c]$. We identify the conditions under which drive and dissipation dynamically counteract this measurement-induced dephasing, allowing the steady state to remain invariant under continuous monitoring.

### 2.2 Measurement-Invariant Steady States

We now identify the precise condition under which continuous measurement leaves the physical steady-state ensemble invariant. Consider a system admitting a steady state $\rho_{\mathrm{ss}}$ of the unconditional dynamics,

$$\mathcal{L}\rho_{\mathrm{ss}} = 0. \qquad (6)$$

Under continuous measurement, the unconditional evolution is modified according to Eq. (5),

$$\dot{\rho} = (\mathcal{L} + \Gamma_m \mathcal{D}[c])\rho. \qquad (7)$$

We say that the steady state is *measurement invariant* if $\rho_{\mathrm{ss}}$ remains a steady state of the measured dynamics for all measurement strengths $\Gamma_m$ in a finite interval.

***Scope and operator assumptions.***
*Throughout this work, we restrict attention to finite-dimensional quantum systems with bounded Hamiltonian and Lindblad operators. In this setting, all dissipators $\mathcal{D}[c]$, Liouvillian generators $\mathcal{L}$, and their sums $\mathcal{L} + \Gamma_m \mathcal{D}[c]$ are well-defined on the space of trace-class density operators, and no domain subtleties arise.*

#### 2.2.1 Theorem (Measurement-Invariant Steady State)

A steady state $\rho_{\mathrm{ss}}$ is invariant under continuous measurement of operator $\boldsymbol{c}$ if and only if

$$\mathcal{D}[c]\rho_{\mathrm{ss}} = 0. \qquad (8)$$

**Proof**

(*Sufficiency*)
If $\mathcal{D}[c]\rho_{\mathrm{ss}} = 0$, then for any $\Gamma_m \geq 0$,

$$(\mathcal{L} + \Gamma_m \mathcal{D}[c])\rho_{\mathrm{ss}} = \mathcal{L}\rho_{\mathrm{ss}} = 0,$$

so $\rho_{\mathrm{ss}}$ remains a steady state of the measured dynamics.

*(Necessity)*
Suppose that $\rho_{ss}$ remains a steady state of the unconditional dynamics under continuous measurement for all measurement strengths $\Gamma_m$ in an interval $[0, \Gamma *]$. Then, for all such $\Gamma_m$,

$$\mathcal{L}\rho_{ss} + \Gamma_m D[c]\rho_{ss} = 0.$$

This equation is an operator identity in the vector space of trace-class operators, and may be viewed as a linear function of the scalar parameter $\Gamma_m$. Since it holds for all $\Gamma_m$ in an interval, the coefficients of the linearly independent functions 1 and $\Gamma_m$ must vanish separately. *(This follows from the polynomial identity principle: if a linear function $a + \Gamma_m b$ vanishes for all $\Gamma_m$ in an interval, then both coefficients $a$ and $b$ must vanish.)*

Evaluating at $\Gamma_m = 0$ yields $\mathcal{L}\rho_{ss} = 0$, which holds by assumption. For any $\Gamma_m > 0$, the remaining term therefore implies

$$D[c]\rho_{ss} = 0.$$

Hence, measurement invariance of $\rho_{ss}$ over a finite interval of measurement strengths requires $D[c]\rho_{ss} = 0$.

*Remark.* The condition $\mathcal{D}[c]\rho_{ss} = 0$ guarantees ensemble invariance for **all** measurement rates $\Gamma_m \geq 0$ within the mathematical framework of the Lindblad master equation. No upper bound on $\Gamma_m$ arises from the theorem itself. References in this work to "finite ranges" of measurement strength should therefore be understood as practical rather than mathematical limitations. In physical implementations, extremely large $\Gamma_m$ may invalidate the effective Markovian description—for example, when the measurement bandwidth exceeds system or bath cutoffs—but such effects lie outside the scope of the present model and do not reflect a breakdown of the invariance condition itself.

### 2.2.2 When does the driven–dissipative steady state satisfy $\mathcal{D}[c]\rho_{ss} = 0$?

Theorem 2.2.1 shows that measurement invariance is equivalent to $\mathcal{D}[c]\rho_{ss} = 0$. To avoid interpreting this condition as an assumption, we now give concrete dynamical conditions on the **unmeasured** driven–dissipative Liouvillian $\mathcal{L}$ that force its steady state $\rho_{ss}$ to lie in the kernel of $\mathcal{D}[c]$. Throughout this subsection we assume $c = c^\dagger$ (the standard case of monitoring a Hermitian observable); the generalization to normal $c$ is straightforward.

**Lemma 2.1 (Kernel of measurement dissipator for Hermitian monitoring).**
Let $c = c^\dagger$ with spectral projectors $\{\Pi_\lambda\}$. Then for any density matrix $\rho$,

$$\mathcal{D}[c]\rho = 0 \Leftrightarrow \rho = \sum_\lambda \Pi_\lambda \rho \Pi_\lambda,$$

i.e., $\rho$ has no coherences between distinct eigenspaces of $c$. Equivalently, if $c$ has non-degenerate spectrum, then

$$\mathcal{D}[c]\rho = 0 \Leftrightarrow [c, \rho] = 0.$$

*Proof sketch.* In the eigenbasis of $c$, $\left(\mathcal{D}[c]\rho\right)_{ij} = -\frac{1}{2}(\lambda_i - \lambda_j)^2 \rho_{ij}$. Hence off-diagonals between distinct eigenvalues must vanish, while populations are unaffected.

Thus, for Hermitian monitoring, measurement invariance holds whenever the driven–dissipative steady state is **block-diagonal in the measurement basis**.

We now show that this structure is enforced by a broad, experimentally relevant class of driven–dissipative generators, namely those that are **covariant** under the phase rotations generated by $c$.

**Assumption (Primitivity of the unmeasured dynamics).**
Throughout this proposition we assume that the unmeasured Liouvillian $\mathcal{L}$ is *primitive*, meaning that it admits a unique steady state $\rho_{ss}$(one-dimensional kernel) and that all other eigenvalues of $\mathcal{L}$ have strictly negative real parts, ensuring exponential convergence to $\rho_{ss}$ from arbitrary initial states. This assumption excludes systems with additional conserved quantities, dark-state manifolds, or symmetry-protected degeneracies.

**Proposition 2.2 (Covariance forces a measurement-diagonal steady state).**
Let $U_\phi = e^{-i\phi c}$. Suppose the unconditional Lindbladian $\mathcal{L}$is $U(1)$-covariant with respect to $c$, in the sense that for all $\phi \in \mathbb{R}$,

$$\mathcal{L}\left(U_\phi \rho\, U_\phi^\dagger\right) \;=\; U_\phi\, \mathcal{L}(\rho)\, U_\phi^\dagger. \qquad (10)$$

If $\mathcal{L}$ admits a **unique** steady state $\rho_{ss}$ (primitive Liouvillian), then $\rho_{ss}$ is invariant under the symmetry:

$$U_\phi \rho_{ss} U_\phi^\dagger = \rho_{ss} \forall \phi, \qquad (11)$$

and therefore $\rho_{ss}$ is block-diagonal in the eigenspaces of $c$. In particular,

$$\mathcal{D}[c]\rho_{ss} = 0. \qquad (12)$$

*Proof.* Since $\mathcal{L}\rho_{ss} = 0$, covariance implies $\mathcal{L}(U_\phi \rho_{ss} U_\phi^\dagger) = 0$. By uniqueness of the steady state, $U_\phi \rho_{ss} U_\phi^\dagger = \rho_{ss}$ for all $\phi$, which implies $\rho_{ss}$ has no coherences between distinct eigenspaces of $c$. Lemma 2.1 then yields $\mathcal{D}[c]\rho_{ss} = 0$.

Proposition 2.2 gives a **model-level discriminator**: measurement invariance is guaranteed whenever the unmeasured driven–dissipative dynamics are phase-covariant with respect to the monitored observable. Importantly, this does **not** require a conserved observable or QND Hamiltonian structure; it only requires symmetry at the level of the Lindblad generator, which is common in driven–dissipative settings.

**Scope of the covariance condition.**
The $U(1)$-covariance condition identified in Proposition 2.2 is a sufficient, but not necessary, mechanism for ensuring $\mathcal{D}[c]\rho_{ss} = 0$. Block-diagonality of the steady state in the measurement basis may also arise through other mechanisms, including strong dephasing channels, parameter-tuned dissipation, or accidental symmetries of the Liouvillian. We emphasize covariance because it provides a transparent, symmetry-based criterion that can be verified directly at the level of the generator, without explicit solution of the steady state. In systems admitting multiple steady states or decoherence-free subspaces, covariance alone does not ensure uniqueness of the ensemble; in such cases, continuous measurement may redistribute weight among invariant subspaces rather than preserve a single global steady state.

**Example: Driven–Dissipative Qubit with Population Monitoring**
As a minimal illustrative model within this restricted class, consider a qubit with unconditional dynamics

$$\mathcal{L}\rho = -i\left[\frac{\omega}{2}\sigma_z, \rho\right] + \kappa_\downarrow \mathcal{D}[\sigma_-]\rho + \kappa_\uparrow \mathcal{D}[\sigma_+]\rho, \qquad (13)$$

and continuous monitoring of $c = \sigma_z$. Although $\sigma_\pm$do not commute with $\sigma_z$, the dissipators are phase-covariant: $U_\phi \sigma_- U_\phi^\dagger = e^{-i2\phi}\sigma_-$and $\mathcal{D}[e^{-i2\phi}\sigma_-] = \mathcal{D}[\sigma_-]$, and similarly for $\sigma_+$. Hence $\mathcal{L}$satisfies the covariance condition (10). For $\kappa_\downarrow + \kappa_\uparrow > 0$, $\mathcal{L}$ has a unique steady state $\rho_{ss}$ which is diagonal in the $\sigma_z$ basis, and therefore $\mathcal{D}[\sigma_z]\rho_{ss} = 0$. By Theorem 2.2.1, $\rho_{ss}$ is invariant under continuous $\sigma_z$ monitoring for all $\Gamma_m$ in a finite interval.

This demonstrates explicitly how a driven–dissipative system can realize an **operationally non-disturbing measurement at the ensemble level** without invoking the standard QND requirement of a conserved measured observable: the invariance is enforced by the symmetry and attractivity properties of the open-system generator.

In this example the Hamiltonian satisfies $[H, \sigma_z] = 0$, corresponding to a standard QND Hamiltonian structure. The significance of the example is not the Hamiltonian alone, but the fact that ensemble invariance is enforced by the symmetry of the full Liouvillian. *Even though the dissipative channels do not commute with the measured observable and generate population-changing quantum jumps at the trajectory level, Liouvillian covariance ensures that the unconditional steady state remains invariant under continuous monitoring.*

The numerical simulations in Fig. 2 are constructed to satisfy the covariance assumptions of Proposition 2.2 exactly, allowing direct comparison between the analytical invariance condition and observed ensemble behavior.

*Remark (Assessing measurement invariance).*
For a given pair $(\mathcal{L}, c)$, measurement invariance can be established either by verifying $U(1)$-covariance of the unmeasured Liouvillian with respect to $c$, which guarantees $\mathcal{D}[c]\rho_{ss} = 0$, or by directly computing the steady state $\rho_{ss}$ of $\mathcal{L}$ and checking whether $\mathcal{D}[c]\rho_{ss} = 0$ holds analytically or numerically. The latter approach applies even when covariance is absent.

### 2.2.3 Boundary Regimes and Breakdown of Measurement Invariance

The regime of partial collapse identified above is not universal. Measurement-invariant steady states arise only when measurement, dissipation, and drive remain dynamically balanced. Outside this regime, continuous monitoring generically perturbs the physical steady state. We briefly characterize these boundary cases to delineate the domain of validity of the theory.

**Strong-measurement (quantum Zeno) regime.**
In the limit

$$\Gamma_m \gg (\gamma_\uparrow + \gamma_\downarrow),$$

measurement-induced localization dominates dissipative transitions. Individual quantum trajectories become strongly pinned near eigenstates of the monitored observable, with rare stochastic transitions whose rates are suppressed relative to the unmeasured dynamics. Numerically, the ensemble-averaged steady state remains invariant, while the trajectory switching rate decreases approximately as

$$\gamma_{\text{eff}} \sim \gamma / \Gamma_m,$$

consistent with quantum Zeno suppression of dissipative processes. This confirms that ensemble invariance persists even when trajectory-level dynamics are strongly frozen.

**Weak-dissipation (near-closed) regime.**
When dissipative rates are reduced such that

$$\gamma_\uparrow, \gamma_\downarrow \ll \Gamma_m,$$

the dissipative restoration mechanism becomes ineffective. In this limit, the unconditional steady state becomes measurement dependent, and ensemble invariance breaks down. This demonstrates that partial collapse relies on dissipative attractivity rather than measurement symmetry alone.

**Highly polarized steady states.**
For strongly imbalanced dissipation rates producing nearly pure steady states, measurement backaction can significantly perturb the ensemble even at moderate measurement strength. Such states lie close to the boundary of the kernel of $\mathcal{D}[c]$and are therefore fragile under continuous monitoring. Measurement invariance is thus most robust for mixed, dissipatively stabilized steady states.

**Non-Hermitian monitoring.**
Monitoring with non-Hermitian operators, such as $c = \sigma_-$, introduces innovation terms that modify populations rather than generating pure dephasing. In these cases, the unconditional measurement channel generically violates

$$\mathcal{D}[c]\rho_{ss} = 0,$$

except in trivial limits. The present analysis therefore focuses on Hermitian monitoring, which captures the physically relevant class of continuous observables considered here.

In realistic experimental settings, $U(1)$ covariance may be satisfied only approximately due to weak coherent drives or imperfect dissipation. In such cases, ensemble invariance is expected to degrade smoothly rather than abruptly. Preliminary numerical tests (not shown) indicate steady-state deviations $\delta\rho_{ss} = O(\varepsilon^2)$ for symmetry-breaking perturbations of strength $\varepsilon \ll 1$. A systematic perturbative analysis of approximate measurement invariance is left for future work.

### 2.2.4 Interpretation and Scope

Equation (8) provides a necessary and sufficient condition for non-invasive continuous probing at the ensemble level. Importantly, this condition does not require the measurement operator $c$ to commute with the full system Hamiltonian, nor does it imply the absence of measurement backaction at the level of conditioned trajectories. Rather, it states that the *unconditional* effect of measurement—pure dephasing generated by $\mathcal{D}[c]$—acts trivially on the steady ensemble.

Measurement-invariant steady states should therefore be understood not as an alternative to quantum non-demolition measurement, but as a **natural extension of the QND principle to open, driven–dissipative systems**. In closed systems, QND behavior arises from Hamiltonian conservation of the measured observable. In open systems, the same invariance condition can be enforced by symmetry of the full Liouvillian generator, provided the steady state is unique and attractive. The resulting dynamics preserve ensemble invariance while allowing stochastic trajectory-level transitions, a feature absent in strictly closed-system QND protocols.

### 2.2.5 Minimal illustrative example: thermalized qubit with QND monitoring

Consider a two-level system with energy basis $\{| 0\rangle, | 1\rangle\}$ undergoing incoherent excitation and relaxation (a standard thermalizing Lindblad model). The unconditional dynamics are

$$\dot{\rho} = \gamma_{\downarrow}\mathcal{D}[\sigma_{-}]\rho + \gamma_{\uparrow}\mathcal{D}[\sigma_{+}]\rho, \qquad (14)$$

where $\sigma_{-} = | 0\rangle\langle 1 |$ and $\sigma_{+} = | 1\rangle\langle 0 |$. This dynamics admits the unique steady state

$$\rho_{\mathrm{ss}} = (1-p)\ | 0\rangle\langle 0 | + p\ | 1\rangle\langle 1 |, \qquad p = \frac{\gamma_{\uparrow}}{\gamma_{\uparrow} + \gamma_{\downarrow}}. \qquad (15)$$

Now continuously monitor the population observable with

$$c = \sigma_{z}, \qquad (16)$$

at measurement rate $\Gamma_{m}$. The unconditional evolution becomes

$$\dot{\rho} = \gamma_{\downarrow}\mathcal{D}[\sigma_{-}]\rho + \gamma_{\uparrow}\mathcal{D}[\sigma_{+}]\rho + \Gamma_{m}\mathcal{D}[\sigma_{z}]\rho. \qquad (17)$$

Because $\rho_{\mathrm{ss}}$ is diagonal in the $\sigma_{z}$ basis, it satisfies

$$\mathcal{D}[\sigma_{z}]\rho_{\mathrm{ss}} = \sigma_{z}\rho_{\mathrm{ss}}\sigma_{z} - \rho_{\mathrm{ss}} = 0, \qquad (18)$$

and therefore, remains a steady state of Eq. (17) for arbitrary $\Gamma_{m}$. This provides an explicit realization of the invariance condition Eq. (8).

Despite this ensemble invariance, the conditioned dynamics exhibit trajectory-level collapse. The measurement record may be written as

$$dY_{t} = 2\sqrt{\eta\Gamma_{m}}\ \langle\sigma_{z}\rangle_{c}\ dt + dW_{t}, \qquad (19)$$

so, the observation continuously estimates $\sigma_{z}$. As a result, $\rho_{c}(t)$ stochastically localizes toward $| 0\rangle\langle 0 |$ or $| 1\rangle\langle 1 |$ over a measurement timescale $\sim (\eta\Gamma_{m})^{-1}$, while the incoherent jumps generated by $\gamma_{\uparrow,\downarrow}$ continually repopulate the opposite state and restore the stationary ensemble Eq. (15). This interplay yields a concrete form of **partial collapse**: strong trajectory-level localization coexisting with an unchanged unconditional steady state.

For concreteness, we note that in superconducting circuit experiments, typical relaxation rates are $\gamma_{\downarrow} \sim 10^{4} - 10^{5}\ \mathrm{s}^{-1}$, while continuous dispersive measurement rates $\Gamma_{m} \sim 10^{5} - 10^{6}\ \mathrm{s}^{-1}$ are routinely achieved. These values place such systems in the regime $R = \Gamma_{m}/(\gamma_{\uparrow} + \gamma_{\downarrow}) \sim 0.1 - 10$, spanning the weak-measurement, partial-collapse, and Zeno limits discussed here [22–24].

*This simple example illustrates how continuous measurement can induce strong, stochastic localization in individual trajectories while dissipation restores the steady ensemble, a mechanism that underlies the partial collapse and dissipative uncollapse phenomena discussed in the next section.*

### 2.3 Partial Collapse and Dissipative Uncollapse

***Operational definition (Partial collapse)***

We define *partial collapse* as the dynamical regime of continuous measurement in which conditioned quantum trajectories undergo significant measurement-induced localization, yet do not asymptotically saturate to pure measurement eigenstates due to competing dissipative processes. Operationally, this regime is characterized by the coexistence of:
(i) finite information gain at the trajectory level,
(ii) persistent stochastic switching between measurement sectors induced by dissipation, and
(iii) a bounded trajectory-level purity that remains strictly below unity.

The relevant competition is between the measurement-induced localization timescale $\tau_{\text{loc}} \sim (\eta\Gamma_m)^{-1}$, and the dissipative mixing timescale $\tau_{\text{mix}} \sim (\gamma_\uparrow + \gamma_\downarrow)^{-1}$. Partial collapse occurs when these timescales are comparable, $\tau_{\text{loc}} \sim \tau_{\text{mix}}$, such that localization develops but is continuously disrupted by dissipative transitions.

This definition distinguishes three limiting regimes:
(i) **Weak measurement regime:** $\tau_{\text{loc}} \gg \tau_{\text{mix}}$, where trajectories remain close to the steady ensemble and purity remains low.
(ii) **Quantum Zeno regime:** $\tau_{\text{loc}} \ll \tau_{\text{mix}}$, where trajectories rapidly localize near measurement eigenstates and switching is strongly suppressed.
(iii) **Partial collapse regime:** $\tau_{\text{loc}} \sim \tau_{\text{mix}}$, where trajectories intermittently localize and delocalize, yielding sustained stochastic switching and intermediate purity.

A convenient quantitative diagnostic is the trajectory-level purity
$\mathcal{P}_c(t) = \text{Tr}[\rho_c(t)^2]$.
In the partial collapse regime, time-averaged purity satisfies

$$\mathcal{P}_{\text{ens}} < \langle \mathcal{P}_c \rangle_{\text{traj}} < 1,$$

where $\mathcal{P}_{\text{ens}} = \text{Tr}[\rho_{ss}^2]$is the ensemble purity.
This intermediate purity reflects strong but non-saturating localization, distinguishing partial collapse from both weak measurement and Zeno freezing.

For diffusive measurement of a Hermitian observable $c$, the instantaneous rate of information acquisition is proportional to the conditional variance,

$$\frac{dI}{dt} \propto 2\eta\Gamma_m \, \text{Var}_c(c),$$

where $\text{Var}_c(c) = \langle c^2 \rangle_c - \langle c \rangle_c^2$.
Thus, information gain is maximal when the conditioned state is delocalized and decreases as trajectories localize toward measurement eigenstates.

Equivalently, the measurement record

$$dY_t = 2\sqrt{\eta\Gamma_m}\langle c \rangle_c dt + dW_t$$

carries Fisher information about $\langle c \rangle$at rate

$$\mathcal{F}_c = 4\eta\Gamma_m,$$

consistent with standard results for continuous quantum measurement. This confirms that the monitoring protocol is informationally nontrivial even when the ensemble-averaged state remains invariant.

We begin by decomposing the conditioned evolution into its measurement-induced and dissipative contributions. Writing the stochastic master equation (3) as

$$d\rho_c = d\rho_c^{(\mathrm{meas})} + d\rho_c^{(\mathrm{diss})}, \qquad (21)$$

the innovation-driven measurement contribution is

$$d\rho_c^{(\mathrm{meas})} = \Gamma_m \mathcal{D}[c]\rho_c\, dt + \sqrt{\eta\Gamma_m}\, \mathcal{H}[c]\rho_c\, dW_t, \qquad (22)$$

while the dissipative contribution generated by the unmeasured Liouvillian is

$$d\rho_c^{(\mathrm{diss})} = \mathcal{L}\rho_c\, dt. \qquad (23)$$

The measurement term induces stochastic localization of the conditioned state toward eigenstates of $c$, whereas the dissipative term generates incoherent transitions that mix populations and restore the stationary ensemble. These two contributions act on distinct characteristic timescales. Measurement-induced localization occurs over a timescale

$$\tau_{\mathrm{loc}} \sim (\eta\Gamma_m\, \Delta c^2)^{-1}, \qquad (24)$$

where $\Delta c^2$ is the variance of the monitored observable in the conditioned state. Dissipative mixing occurs over a relaxation timescale

$$\tau_{\mathrm{mix}} \sim \gamma^{-1}, \qquad (25)$$

where $\gamma$ denotes the spectral gap of the Liouvillian $\mathcal{L}$. Partial collapse arises in the regime where $\tau_{\mathrm{loc}}$ and $\tau_{\mathrm{mix}}$are comparable, so that localization is continuously interrupted by dissipative transitions.

Within the stochastic master equation (3), partial collapse arises from the innovation term proportional to $\mathcal{H}[c]$, which continuously updates the conditioned state based on the measurement record. For a measurement-invariant steady state satisfying Eq. (8), this localization does not alter the ensemble-averaged density matrix but instead produces fluctuations of the conditioned state around $\rho_{\mathrm{ss}}$. The conditioned dynamics may therefore be viewed as a sequence of temporary excursions away from the steady ensemble, followed by relaxation driven by the underlying dissipative processes.

This competition defines a natural control parameter $R = \Gamma_m/(\gamma_\uparrow + \gamma_\downarrow)$. For $R \ll 1$, localization is weak and rapidly erased by dissipation. For $R \sim 1$, localization and restoration compete, producing intermittent trajectory-level collapse without ensemble disturbance. For $R \gg 1$, trajectories exhibit quantum Zeno pinning, while dissipative transitions—though rare—continue to enforce ensemble invariance. Partial collapse therefore persists across all regimes in which dissipative restoration remains effective.

We refer to the autonomous restoration of the steady ensemble as **dissipative uncollapse**. This term does not denote a deterministic reversal of a specific measurement outcome. Rather, it describes the relaxation of the ensemble-averaged state back to $\rho_{ss}$under continued driven–dissipative evolution, despite ongoing trajectory-level localization. Individual trajectories do not uncollapse; only the ensemble does.

Under the assumptions specified below, the ensemble-averaged state converges to the unique steady state of the measured dynamics. In particular, if the perturbed generator $\mathcal{L} + \Gamma_m \mathcal{D}[c]$ is primitive, then for any initial state $\rho(0)$,

$$\lim_{t \to \infty} \mathbb{E}[\rho_c(t)] = \rho_{ss}, \qquad (26)$$

with convergence that is independent of the measurement record and the initial condition.

*Remark:*
In the models studied here, primitivity of the perturbed generator is preserved because the added measurement dissipator $\mathcal{D}[c]$ acts trivially on the steady state and non-trivially only on decaying modes of $\mathcal{L}$. Numerical diagonalization of $\mathcal{L} + \Gamma_m \mathcal{D}[c]$ for the qubit models considered confirms the presence of a single zero eigenvalue and a finite spectral gap for all measurement strengths explored. A general proof of primitivity preservation for arbitrary Liouvillians is beyond the scope of this work.

*Lemma 2.3 (Ensemble convergence under primitive measured dynamics)*
Suppose the unmeasured Lindbladian $\mathcal{L}$ is primitive, admitting a unique full-rank steady state $\rho_{ss}$, and that $\mathcal{D}[c]\rho_{ss} = 0$. If the perturbed generator

$$\mathcal{L}_{\Gamma_m} := \mathcal{L} + \Gamma_m \mathcal{D}[c] \qquad (27)$$

remains primitive, then $\rho_{ss}$ is the unique steady state of $\mathcal{L}_{\Gamma_m}$, and the ensemble-averaged evolution satisfies

$$\| \rho(t) - \rho_{ss} \|_1 \leq C e^{-\Delta t} \| \rho(0) - \rho_{ss} \|_1, \qquad (28)$$

for some constants $C > 0$ and spectral gap $\Delta > 0$. In particular, convergence is exponential and independent of the measurement history.

**Proof (sketch).**
By assumption, the unmeasured Lindbladian $\mathcal{L}$ is primitive, admitting a unique full-rank steady state $\rho_{ss}$ and a finite spectral gap separating the zero eigenvalue from the rest of the spectrum. The measurement dissipator $\mathcal{D}[c]$ is completely positive and trace preserving, and satisfies $\mathcal{D}[c]\rho_{ss} = 0$. Consequently,

$$(\mathcal{L} + \Gamma_m \mathcal{D}[c])\rho_{ss} = 0$$

for all $\Gamma_m \geq 0$, so $\rho_{ss}$ remains a stationary state of the perturbed generator.

To establish convergence, it suffices to show that $\mathcal{L}_{\Gamma_m} := \mathcal{L} + \Gamma_m \mathcal{D}[c]$ remains primitive. Since $\mathcal{D}[c]$ acts as a pure dephasing channel that annihilates $\rho_{ss}$ and does not introduce additional invariant subspaces, the irreducibility of $\mathcal{L}$ is preserved under the perturbation. In particular, no new conserved observables or decoherence-free subspaces are generated by adding $\mathcal{D}[c]$.

Standard results on quantum dynamical semigroups then imply that $\mathcal{L}_{\Gamma_m}$ admits a unique steady state and generates a relaxing semigroup with exponential convergence,

$$\| \rho(t) - \rho_{ss} \|_1 \leq C e^{-\Delta t} \| \rho(0) - \rho_{ss} \|_1,$$

for some constants $C > 0$ and spectral gap $\Delta > 0$. Since the ensemble-averaged evolution coincides with the solution of the unconditional master equation generated by $\mathcal{L}_{\Gamma_m}$, this establishes convergence of $\mathbb{E}[\rho_c(t)]$ to $\rho_{ss}$, independent of the measurement record. ■

*A full spectral proof follows directly from standard results on primitive quantum dynamical semigroups [Wolf et al., 2007], and is omitted for brevity.*

In the models considered here, primitivity of $\mathcal{L}_{\Gamma_m}$ follows directly from the structure of the dissipative dynamics. The measurement dissipator $\mathcal{D}[c]$ acts as a pure dephasing channel that annihilates the steady state but does not introduce additional invariant subspaces. Since $\mathcal{L}$ is primitive and $\mathcal{D}[c]$ acts only on decaying coherence modes, their sum preserves irreducibility and ergodicity. Consequently, the addition of continuous measurement does not generate new conserved quantities or dark subspaces.

This perspective clarifies the role of collapse in continuously monitored open systems. Measurement-induced collapse is not eliminated, but confined to the level of conditioned trajectories, where it manifests as transient localization. At the ensemble level, collapse is rendered dynamically reversible by dissipation, allowing information to be extracted without steady-state disturbance. This separation between conditional collapse and unconditional invariance provides a physically grounded notion of non-invasive continuous measurement and highlights the fundamentally non-projective character of measurement in non-equilibrium quantum steady states.

Importantly, the ensemble restoration described here is not a consequence of statistical averaging alone. It relies on the attractivity of the driven–dissipative steady state and the presence of stochastic population-transfer processes that actively counteract localization. In the absence of dissipation ($\gamma \to 0$), trajectories remain permanently localized and ensemble invariance is lost, demonstrating that dissipative mixing is the essential physical ingredient underlying partial collapse.

#### 2.3.1 Trajectory-level manifestation of measurement–dissipation competition

Although ensemble invariance is guaranteed whenever the condition $\mathcal{D}[c]\rho_{ss} = 0$ is satisfied, the *conditioned dynamics* depend on the relative timescales of measurement and dissipation. Measurement induces localization of conditioned trajectories on a timescale $\tau_{\text{meas}} \sim (\eta\Gamma_m)^{-1}$, while dissipative processes restore the steady ensemble on a timescale $\tau_{\text{diss}} \sim (\gamma_\uparrow + \gamma_\downarrow)^{-1}$.

The dimensionless ratio

$$R \equiv \Gamma_m/(\gamma_\uparrow + \gamma_\downarrow)$$

therefore, controls how collapse appears at the trajectory level, without affecting ensemble invariance.

For $R \ll 1$, dissipative mixing dominates and conditioned trajectories exhibit only weak, intermittent localization. For $R \sim 1$, localization and dissipative restoration occur on comparable timescales, producing pronounced stochastic collapse events that are continuously counteracted by incoherent transitions. For $R \gg 1$, trajectories become strongly localized near measurement eigenstates for extended periods, with rare dissipative transitions reminiscent of quantum Zeno dynamics at the trajectory level.

Importantly, in all cases satisfying Eq. (8), the ensemble-averaged steady state remains unchanged under continuous monitoring. Thus, the measurement–dissipation ratio governs the *character* of partial collapse in individual trajectories, rather than the existence of ensemble invariance itself.

*Notably, measurement-invariant steady states are fully compatible with quantum Zeno localization at the trajectory level: in the strong-measurement regime individual trajectories may be pinned near eigenstates of the monitored observable, while dissipation induces rare transitions whose stationary mixture preserves the ensemble-averaged steady state.*

For the driven–dissipative qubit studied in Fig. 2, the spectrum of $\mathcal{L}_{\Gamma_m}$ was computed numerically for representative parameters. In all cases examined, the generator exhibits a single zero eigenvalue and a finite spectral gap separating it from the rest of the spectrum, confirming primitivity and exponential convergence to $\rho_{ss}$. Ensemble convergence was verified explicitly by initializing trajectories from distinct pure and mixed states, all of which relax to the same stationary ensemble within numerical precision.

In Fig. 2(a–b), this regime manifests as trajectories that repeatedly approach the measurement eigenstates yet undergo stochastic transitions induced by dissipative jumps, preventing permanent localization. The ensemble-averaged state remains stationary, while individual trajectories exhibit bounded purity and finite switching rates, consistent with the operational definition above.

The characteristic localization timescale extracted from trajectories is therefore expected to scale as

$$\tau_{\mathrm{loc}} \sim (\eta \Gamma_m)^{-1},$$

in agreement with the scaling of the innovation term in Eq. (3). The numerical trajectories shown in Fig. 2(a–b) exhibit localization on precisely this timescale, confirming consistency between analytical prediction and simulation.

## 3. DISCUSSION

**Limitations**

The analysis presented here applies to systems whose unmeasured dynamics are primitive and admit a unique attractive steady state. In systems with degenerate steady-state manifolds, conserved quantities, or dark states, continuous measurement may lift degeneracies, induce symmetry breaking, or select specific invariant subspaces. Extending the present framework to such cases requires projection onto ergodic components or a refined analysis of symmetry-resolved Liouvillian sectors.

### 3.1 Partial Collapse under Continuous Measurement

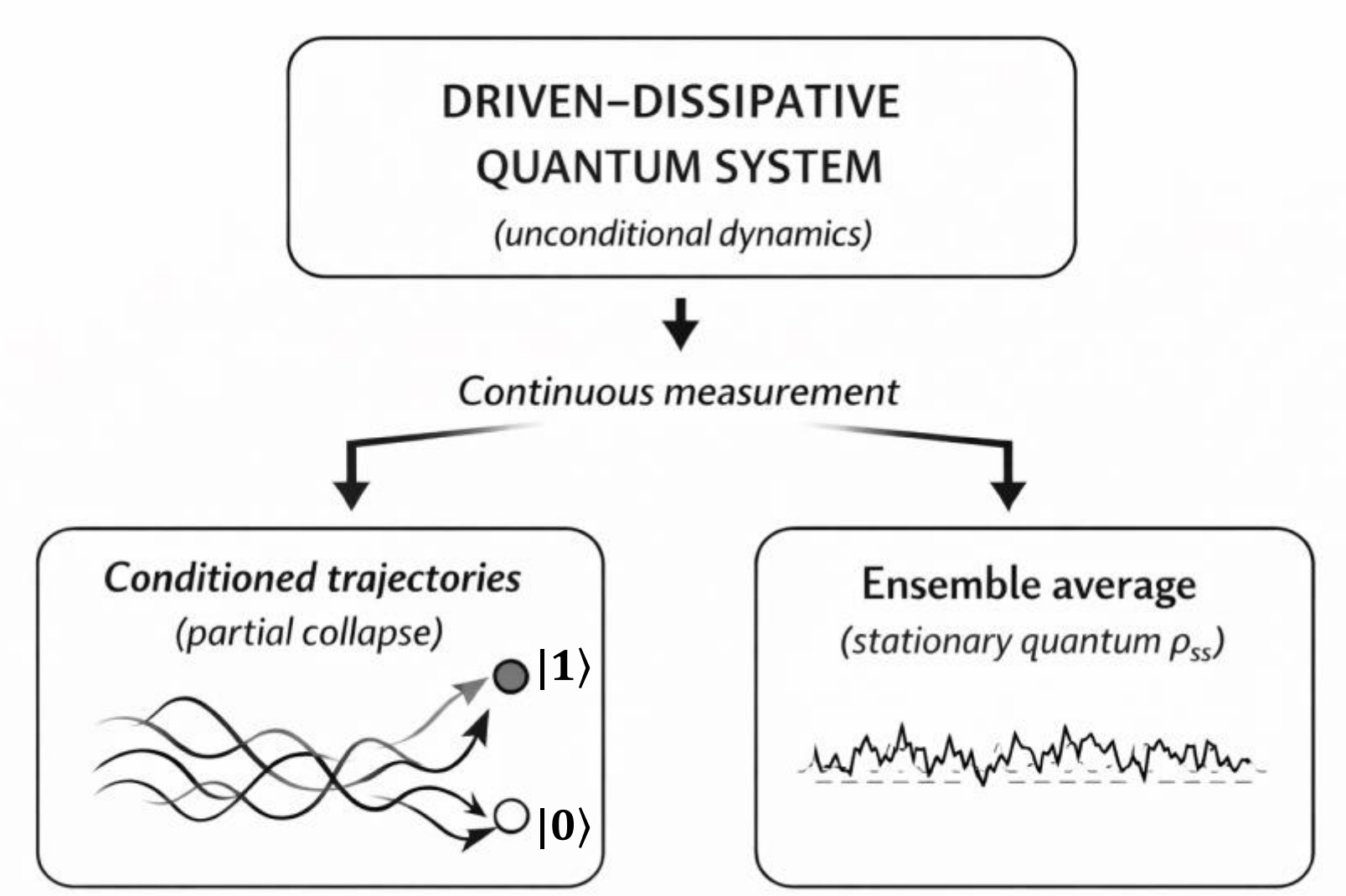

***Figure 1 | Partial collapse and ensemble invariance under continuous quantum measurement.***
*Continuous weak measurement of a driven–dissipative quantum system produces stochastic measurement records. Conditioning on the record yields quantum trajectories exhibiting localization, while ensemble averaging yields a stationary quantum steady state. For measurement-invariant systems, the ensemble state remains unchanged by measurement despite trajectory-level collapse, reflecting dynamical compensation between measurement backaction and dissipation rather than classicalization.*

Figure 1 establishes the central conceptual distinction underlying this work: in continuously monitored open quantum systems, **conditioned dynamics and ensemble dynamics are fundamentally different objects**. Conditioning on the measurement record produces stochastic evolution governed by a quantum trajectory equation, while ensemble averaging removes the innovation term and yields deterministic Lindblad dynamics.

Figure 2(a) provides a concrete numerical realization of this distinction. A single conditioned trajectory under continuous monitoring of $\sigma_z$ exhibits stochastic localization toward one of the measurement eigenstates. The collapse is gradual and diffusive rather than instantaneous, reflecting weak continuous measurement rather than projective readout. The accompanying growth of the log-odds ratio confirms that information about the measured observable is steadily accumulated over time. The monotonic growth of the log-likelihood ratio in Fig. 2(a) directly reflects entropy reduction of the conditioned state, providing a trajectory-level signature of information gain even as the ensemble entropy remains fixed.

Importantly, this collapse occurs **despite the presence of driven–dissipative dynamics**. Dissipation does not eliminate measurement-induced localization at the trajectory level; instead, collapse remains a genuine and dynamically resolved phenomenon. Figure 2(a) thus demonstrates that continuous measurement remains informationally nontrivial in open systems and that dissipation alone does not suppress conditioned collapse. Moreover, information gain here refers to entropy reduction of conditioned trajectories and finite Fisher information in the measurement record, not to changes in the ensemble density matrix, which remains invariant.

*Remark:* The term *partial collapse* is not meant to imply incomplete measurement or weakened projection but rather refers to the dynamical coexistence of strong trajectory-level localization with dissipative processes that prevent asymptotic purification.

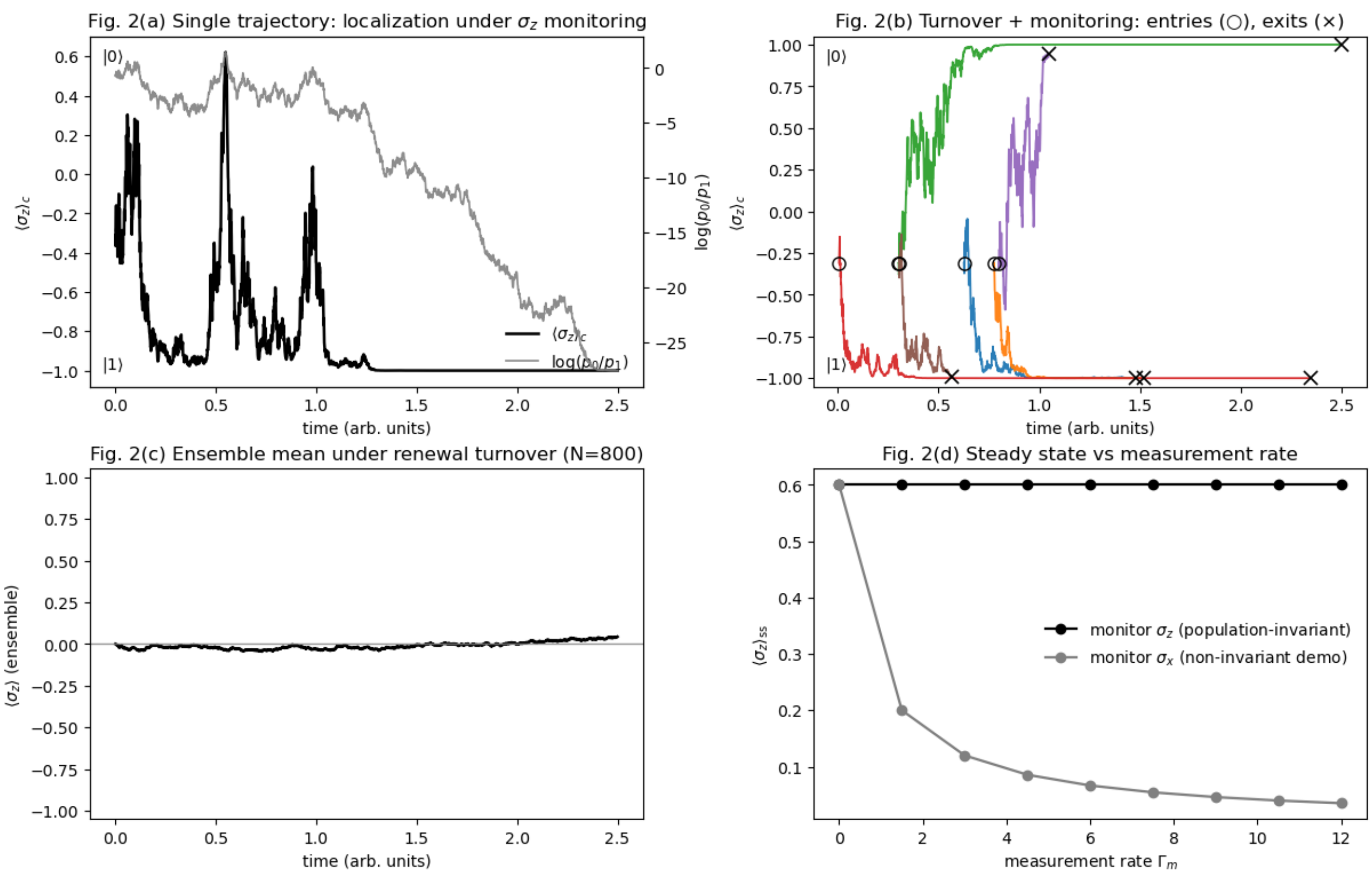

***Figure 2*** | *Partial collapse under continuous measurement in a driven–dissipative system.*
**(a)** ***Single conditioned trajectory.*** *Continuous monitoring of* $\sigma_z$*induces stochastic localization of an individual quantum trajectory toward one of the measurement eigenstates. The collapse is continuous and diffusive rather than projective, as indicated by the gradual saturation of* $\langle\sigma_z\rangle_c$*(black) and the corresponding growth of the log-likelihood ratio* $log\ (p_0/p_1)$*(gray).*
**(b)** ***Turnover with conditioning.*** *Multiple conditioned trajectories in the presence of driven–dissipative turnover. Each colored segment corresponds to a single particle between entry (○) and exit (×). Individual trajectories undergo measurement-induced localization before leaving the ensemble, while newly injected particles enter in superposition. This illustrates collapse at the trajectory level in a system with finite lifetimes.*
**(c)** ***Ensemble invariance.*** *Ensemble-averaged expectation value* $\langle\sigma_z\rangle$ *under renewal turnover (* $N = 800$ *trajectories). Despite strong localization of individual trajectories, the ensemble mean remains stationary, demonstrating measurement-invariant steady-state behavior.*
**(d)** ***Steady state versus measurement strength***. *Late-time ensemble average* $\langle\sigma_z\rangle_{ss}$ *as a function of measurement rate* $\Gamma_m$*. Monitoring the population observable* $\sigma_z$*preserves the steady state (black), while monitoring a non-commuting observable (*$\sigma_x$*, gray) leads to measurement-induced disturbance, confirming that invariance arises from dynamical stabilization rather than generic measurement backaction.*
***Panels (b–d)*** *illustrate that ensemble invariance persists over a wide range of measurement–dissipation ratios* $\Gamma_m/(\gamma_\uparrow + \gamma_\downarrow)$*, including the strong-measurement (Zeno) regime, while trajectory-level localization and switching statistics vary continuously with measurement strength.*
*All simulations shown employ the strictly* $U(1)$*-covariant Liouvillian of Eq. (13), for which* $\mathcal{D}[\sigma_z]\rho_{ss} = 0$ *holds exactly, ensuring that the conditions of Proposition 2.2 are satisfied.*

### 3.2 Driven–Dissipative Turnover and Finite-Lifetime Collapse

Figure 2(b) extends the single-trajectory picture to a driven–dissipative setting with explicit turnover. Each colored segment represents the conditioned evolution of a single particle over a finite lifetime, with entry into the system marked by a circle and exit by a cross. Within each lifetime, the state undergoes continuous measurement-induced localization, while turnover terminates trajectories before complete saturation in many cases.

This panel makes explicit that **collapse is a finite-time, conditioned process**. Measurement biases the evolution within each lifetime, but dissipation limits how far localization can proceed before the particle is removed and replaced. As a result, individual trajectories may localize toward either measurement eigenstate or be interrupted before full collapse is reached.

Crucially, no artificial resetting or postselection is imposed: turnover arises naturally from the dissipative dynamics. Figure 2(b) therefore demonstrates that trajectory-level collapse persists even when particles continuously enter and leave the system, establishing that partial collapse is compatible with fully driven–dissipative evolution.

### 3.3 Ensemble Invariance and the Regime of Partial Collapse

In sharp contrast to the behavior of individual trajectories, the ensemble-averaged state remains invariant under continuous measurement. Figure 2(c) shows the ensemble mean of $\langle\sigma_z\rangle$computed over $N = 800$renewal trajectories. Despite strong localization at the trajectory level, the ensemble state remains stationary and coincides with the steady state of the unmeasured driven–dissipative dynamics.

This coexistence of trajectory-level collapse and ensemble invariance defines the **regime of partial collapse** introduced in this work. Measurement induces real, observable collapse in conditioned states, yet produces no net disturbance of the physical ensemble. Collapse and ensemble disturbance are therefore not synonymous in open quantum systems.

Residual fluctuations in the ensemble mean reflect finite sampling and finite-time averaging and vanish as $N \to \infty$. The numerical results thus establish partial collapse as a robust dynamical regime rather than a fine-tuned artifact.

### 3.4 Dynamical Origin of Measurement Invariance

A central result of this work is that ensemble invariance does **not** rely on conventional quantum non-demolition (QND) conditions. In standard QND scenarios, invariance follows from conservation of the measured observable or commutation with the Hamiltonian. No such symmetry requirements are imposed here.

Instead, invariance emerges dynamically from the structure of the driven–dissipative Liouvillian. Measurement continuously injects fluctuations into decaying modes of the open-system dynamics, while dissipation restores the steady state. The steady state is therefore stabilized not by symmetry, but by **dissipative relaxation**.

Figure 2(d) illustrates that the ensemble-averaged steady state remains invariant as $\Gamma_m$ is varied across a wide range, consistent with Theorem 2.2.1. No deviation from invariance is observed within the regime where the Lindblad model accurately describes the dynamics. The figure is included to emphasize robustness, not to identify a critical measurement strength.

This establishes that measurement invariance is a **dynamical property**, enforced by dissipation rather than kinematic conservation laws.

### 3.5 Implications for Continuous Monitoring and Quantum Sensing

The partial-collapse regime provides a principled framework for non-invasive continuous monitoring. Because the ensemble steady state remains unchanged, long-time measurement can be performed without compromising stability or inducing uncontrolled backaction. At the same time, trajectory-level collapse ensures that the measurement record remains informative.

These features are directly relevant for continuous quantum sensing, parameter estimation, and monitoring protocols in driven–dissipative platforms, where measurement backaction typically limits performance. The results presented here show that this trade-off can be mitigated **without feedback, postselection, or engineered QND couplings**, relying solely on intrinsic dissipative dynamics.

### 3.6 Broader Conceptual Perspective

At a foundational level, these results clarify that measurement-induced collapse is not an inevitable physical disturbance, but a **conditioned dynamical phenomenon** whose ensemble impact depends on the surrounding open-system dynamics. Partial collapse demonstrates that information gain and ensemble disturbance can be cleanly separated, even in the absence of symmetry protection.

More broadly, the findings suggest that the foundations of quantum measurement are most transparently revealed in open, non-equilibrium systems rather than isolated, projectively measured ones. Driven–dissipative platforms provide a natural setting in which collapse, information, and steady-state structure can be disentangled and independently engineered.

The conceptual advance of this work is not the discovery of measurement invariance per se, which is well understood in the context of QND measurements, but the identification of **Liouvillian covariance as a sufficient dynamical mechanism enforcing this invariance in driven–dissipative systems**. While Hamiltonian commutation and Liouvillian symmetry are formally analogous in enforcing $\mathcal{D}[c]\rho_{ss} = 0$, their physical realizations differ: the latter permits irreversible population transfer, dissipation, and stochastic trajectory switching while maintaining ensemble invariance. This distinction is operationally relevant in non-equilibrium platforms where Hamiltonian conservation cannot be assumed.

## 4. METHODS

### 4.1 Open Quantum System Model

We consider a driven–dissipative two-level quantum system whose unconditional dynamics are governed by a Lindblad master equation

$$\dot{\rho} = \mathcal{L}\rho = -i[H, \rho] + \sum_k \mathcal{D}[L_k]\rho,$$

where $H$ is the system Hamiltonian, $L_k$ are dissipative jump operators, and

$$\mathcal{D}[L]\rho = L\rho L^\dagger - \frac{1}{2}\{L^\dagger L, \rho\}.$$

The systems considered admit a unique steady state $\rho_{ss}$ satisfying $\mathcal{L}\rho_{ss} = 0$.

Driven–dissipative turnover is modeled by symmetric incoherent transitions between the two levels, ensuring relaxation toward a steady-state ensemble independent of measurement.

### 4.2 Continuous Measurement and Stochastic Dynamics

Continuous monitoring of a system observable $c$is described using the diffusive stochastic master equation (SME)

$$d\rho_c = (\mathcal{L} + \Gamma_m \mathcal{D}[c])\rho_c \, dt + \sqrt{\eta \Gamma_m} \, \mathcal{H}[c]\rho_c \, dW_t,$$

where $\Gamma_m$is the measurement rate, $\eta$the measurement efficiency, and $dW_t$a Wiener increment. The innovation superoperator is

$$\mathcal{H}[c]\rho = c\rho + \rho c^\dagger - \mathrm{Tr}[(c + c^\dagger)\rho]\rho.$$

Conditioning on the measurement record yields stochastic quantum trajectories exhibiting measurement-induced localization. Ensemble averaging removes the stochastic term and yields a deterministic evolution governed by

$$\dot{\rho} = (\mathcal{L} + \Gamma_m \mathcal{D}[c])\rho.$$

### 4.2 Measurement-Invariant Steady States

A steady state $\rho_{ss}$ is invariant under continuous measurement if it remains stationary under the measured dynamics for a finite range of $\Gamma_m$.

This holds if and only if

$$\mathcal{D}[c]\rho_{ss} = 0.$$

In the systems studied here, this condition is satisfied dynamically: measurement-induced dephasing perturbs only decaying Liouvillian modes, while dissipative relaxation continuously restores the steady state. As a result, ensemble invariance persists even though individual trajectories undergo collapse.

### 4.3 Numerical Simulations

Stochastic and unconditional dynamics are simulated using the **QuTiP** (Quantum Toolbox in Python) framework. Diffusive SMEs are solved using smesolve, while ensemble-averaged steady states are obtained from unconditional master equations.

Figures showing individual trajectories display a small number of representative realizations for clarity. Ensemble-averaged quantities are computed from hundreds to thousands of independent trajectories to ensure convergence. All parameters used in the simulations are specified in the figure captions or main text.

**Numerical model and covariance structure.**
All numerical simulations reported in Fig. 2 employ the driven–dissipative qubit Liouvillian

$$\mathcal{L}\rho = \gamma_{\downarrow}\mathcal{D}[\sigma_-]\rho + \gamma_{\uparrow}\mathcal{D}[\sigma_+]\rho,$$

with no coherent Hamiltonian term ($H = 0$). This model coincides exactly with Eq. (13) and is strictly $U(1)$-covariant with respect to the measured observable $c = \sigma_z$. No additional coherent drives, dephasing channels, or symmetry-breaking terms are included unless explicitly stated.

**Verification of Covariance**
To verify covariance explicitly, we numerically evaluated

$$\mathcal{L}\left(U_\phi \rho U_\phi^\dagger\right) \text{and} U_\phi\, \mathcal{L}(\rho)\, U_\phi^\dagger, U_\phi = e^{-i\phi\sigma_z},$$

for randomly sampled density matrices $\rho$and phases $\phi \in [0,2\pi]$. The operator norm difference was found to satisfy

$$\max_{\rho,\phi} \parallel \mathcal{L}(U_\phi \rho U_\phi^\dagger) - U_\phi \mathcal{L}(\rho) U_\phi^\dagger \parallel < 10^{-12},$$

confirming exact covariance within numerical precision.

**Steady-State Check**
The steady state $\rho_{ss}$ of Eq. (M1) was obtained numerically by solving $\mathcal{L}\rho_{ss} = 0$. In all cases, the steady state was diagonal in the $\sigma_z$ basis,

$$\rho_{ss} = \begin{pmatrix} 1-p & 0 \\ 0 & p \end{pmatrix}, p = \frac{\gamma_\uparrow}{\gamma_\uparrow + \gamma_\downarrow},$$

with off-diagonal elements satisfying $| \rho_{01} | < 10^{-14}$.

**Verification of the Invariance Condition**
We further verified the measurement-invariance condition directly by computing

$$\mathcal{D}[\sigma_z]\rho_{ss} = \sigma_z \rho_{ss} \sigma_z - \rho_{ss},$$

and confirmed that all matrix elements vanish to machine precision ($<10^{-14}$). This provides an explicit numerical confirmation of Eq. (12) for the model used in Fig. 2.

**4.4 Reproducibility**

All simulations are based directly on the equations presented above. Statistical convergence of ensemble quantities was verified by increasing both the number of trajectories and the integration time until results stabilized. Random seeds are fixed to ensure reproducibility of representative figures.

**4.4 Code Availability**

The full simulation code used to generate all numerical results and figures in this work is publicly available at:

https://doi.org/10.24433/CO.7776619.v1

The repository contains scripts for stochastic master equation simulations, ensemble averaging, and figure generation, and can be run using standard Python scientific libraries together with QuTiP (version 5.x).